\documentclass[a4paper, 10pt, twocolumn]{article}

\usepackage[utf8]{inputenc}         

\usepackage{graphicx}               
\graphicspath{ {./images/} }
\usepackage{tabularx}               
\usepackage{multirow}               
\usepackage{url}                

\usepackage[small,bf]{caption2}     
\usepackage{parskip}
\usepackage{titlesec}
\usepackage{amsmath}                

\titleformat{\section}{\normalfont\large\bfseries}{\thesection}{}{}
\titleformat{\subsection}{\normalfont\large\bfseries}{\thesection}{}{}
\titleformat{\paragraph}{\normalfont\bfseries}{\theparagraph}{}{}
\titlespacing{\section}{0pt}{6pt}{-1pt}
\titlespacing{\subsection}{0pt}{3pt}{-1pt}
\titlespacing{\paragraph}{0pt}{3pt}{-1pt}

\newcolumntype{Y}{>{\centering\arraybackslash}X}    

\begin{document}

\date{}                                         

\title{\vspace{-8mm}\textbf{\large
Impact of spatial auditory navigation on user experience\\
during augmented outdoor navigation tasks}}

\author{
Jan-Niklas Voigt-Antons$^1$, Zhirou Sun$^2$, Maurizio Vergari$^2$,\\ Navid Ashrafi$^1$, Francesco Vona$^1$, and Tanja Koji\'c$^2$\\
$^1$ \emph{\small Immersive Reality Lab, Hamm-Lippstadt University of Applied Sciences,
Germany, Email: jan-niklas.voigt-antons@hshl.de
}\\
$^2$ \emph{\small Quality and Usability Lab, Technische Universit\"at Berlin, Germany} } \maketitle
\thispagestyle{empty}           

\section*{Abstract}
\label{sec:Abstract} The auditory sense of humans is important when it comes to navigation. The importance is especially high in cases when an object of interest is visually partly or fully covered. Interactions with users of technology are mainly focused on the visual domain of navigation tasks. This paper presents the results of a literature review and user study exploring the impact of spatial auditory navigation on user experience during an augmented outdoor navigation task. For the user test, participants used an augmented reality app guiding them to different locations with different digital augmentation. We conclude that the utilization of the auditory sense is yet still underrepresented in augmented reality applications. In the future, more usage scenarios for audio-augmented reality such as navigation will enhance user experience and interaction quality.

\section*{Introduction}
\label{sec:introduction}
In recent years, Augmented Reality (AR) has undergone rapid development  mainly thanks to the rapid technological advances that have taken place on the hardware side, with increasingly high-performance and efficient systems and software with advanced computer vision algorithms. In short, AR technology allows users to see and interact with virtual objects or information in the real world, enhancing their perception and experience of their surroundings. One of the earliest definitions of AR is that of Azuma \cite{Azuma1997-wd}, who defines AR as a system with three main characteristics: i) combination of real and virtual: AR systems superimpose virtual information on the real world, allowing users to perceive both in the same time, ii) real-time interaction: AR systems respond to user input and update virtual information in real-time as the user moves or interacts with the environment, iii) 3D registration: AR systems align virtual objects with the physical world in 3D space, allowing them to be positioned and anchored to specific physical locations. From this definition, there seems to be an apparent inclination to consider AR primarily from a visual point of view, as a graphical overlay that is superimposed on the natural world, and this belief is also confirmed by the many AR applications available on the market today such as Pokemon Go. However, sight is not the only sense we use in everyday life. Indeed, almost all human interactions are multimodal, or rather multisensory, in that they involve using more than one sense. This is why using only sight when interacting with an AR system can limit the user experience, especially in specific tasks such as navigation or exploring known and unknown environments. The auditory sense of humans is essential when it comes to navigation. The importance is exceptionally high in cases where the target object is visually covered wholly or partially. AR overlay may, therefore, not only be visual but must be complemented by adding another element, namely the auditory element. Using audio, especially in tasks such as navigation, offers several advantages: i) greater range: the hearing range is greater than the visual range because audio is not affected by the occlusion that can occur with holograms; ii) multitasking: using audio for navigation allows one to use one's eyesight to search for other information or to do just other things while reaching a position, and iii) reduced visual fatigue: the use of audio allows one to alleviate stress on the eyes and consequently reduce the possibility of eye-strain. For these reasons, the impact of spatial auditory navigation on user experience during an outdoor augmented navigation task and its implications were investigated.

\section*{Related work}
\label{sec:related_work} 
In order to acknowledge the current state of research concerning the use of audio in Augmented Reality experiences, a short literature review was conducted. One of the first works to explore the use of spatial audio in AR experiences was that of Sundareswaran et al. \cite{Sundareswaran2003-qd}, who, as early as 2003, created a 3D audio wearable system that can be used to provide alerts and informational cues to a mobile user in such a manner as to appear to emanate from specific locations in the user's environment. The system, although cumbersome, is one of the first successful experiments to integrate audio within AR experiences. 
The possibility of using audio in AR efficiently was also demonstrated in the study by \cite{Bauerfeind2021-vy} in which users' cognitive load while using a driving simulator was analyzed. Navigation information was presented to the user via a head-up display (HUD) or an AR display. Meanwhile, the driver had to solve a non-driving-related task (NDRT), an auditory and spatial cognitive task. The results showed that while driving with the AR display, participants performed better on the NDRT, indicating a reduced mental load compared to the HUD.

As can be seen, the use cases in which spatial audio shows it's potential the most, are orientation and navigation tasks as in the examples of \cite{Miyakoshi2021-jw, Russell2016-ks} in which different setups are tested including those with spatial audio that perform better during navigation tasks. For instance, \cite{Russell2016-ks} presents a wearable device and infrastructure demonstrating the feasibility of scalable indoor and outdoor Auditory Augmented Reality (AAR). In \cite{Miyakoshi2021-jw}, a novel paradigm was presented, the AudioMaze, in which participants freely explore a room-sized virtual maze while an electroencephalogram (EEG) is recorded synchronized to motion capture.

Furthermore, in the case of navigation, spatial audio is a precious resource for visually impaired users. Two examples are the works by \cite{Blum2012-hd, Katz2012-gr}, who present two smartphone applications that use spatial audio to increase the autonomy of blind people during exploration or navigation tasks. This topic is addressed in even more detail in \cite{Afonso-Jaco2022-vw}, where a review of several studies employing the power of spatial audio virtual Reality for spatial cognition research with blind people is presented. These include studies analyzing simple spatial configurations, architectural navigation, sound reaching, and sound design to improve acceptability. 

Nevertheless, spatial audio can also contribute to the realism of the experience, as in the case of the Augmented Reality Audio Game Audio Legends \cite{Rovithis2019-ay}, in which spatial audio is added to augment the realism of the gestural interactions, resulting in an enhanced immersive game experience. 
The works of \cite{Martens2021-bm, Vazquez-Alvarez2016-wm} instead study how to design AR systems with efficient spatial audio. In the latter, the practical limitations of currently available off-the-shelf hardware are evaluated. Results show that mobile audio augmented reality systems achieve the same resolution as stationary systems. The same authors also conducted a study on simplifying the measurement of user orientation in mobile audio augmented reality applications by comparing the orientation of the body with that of the head. The findings show that orientation with the body provides a good approximation when the object or place to be reached is far away while orientation with the head becomes important when the object to be found becomes closer. In \cite{Vazquez-Alvarez2012-ni}, four different auditory displays in a mobile audio-augmented reality environment (a sound garden) were evaluated. The auditory displays varied using non-speech audio, Earcons as auditory landmarks, and 3D audio spatialization. The goal was to test the user experience of discovery in a purely exploratory environment that included multiple simultaneous sound sources. Results show that spatial audio, together with Earcons, allowed users to explore multiple simultaneous sources and had the added benefit of increasing the level of immersion in the experience. Instead, another study investigated multilevel auditory displays to enable eyes-free mobile interaction with indoor location-based information in non-guided audio-augmented environments.
Martens et al. \cite{Martens2021-bm} evaluated the performance of spatial navigation by seated users in multimodal augmented reality systems. The results of those investigations have implications for the effective use of the auditory component of a multimodal AR system in applications supporting spatial navigation through a physical environment.

\section*{Theoretical implication} 
\label{sec:theoretical_implication}
Spatial aural navigation can improve the user experience during augmented outdoor navigation activities by delivering extra sensory signals that complement or replace visual clues. Visual clues in outdoor situations might be limited or covered by variables such as weather, vegetation, or distance, making it difficult for users to navigate efficiently. In contrast, spatial auditory navigation offers users extra information about their surroundings, such as the direction and distance of items, landmarks, and areas of interest. This information can be communicated by auditory cues such as beeps, tones, or voice directions, which can assist users in orienting themselves and making educated navigation decisions.
Furthermore, spatial auditory navigation can lessen the cognitive burden on users by allowing them to engage with the environment in a more natural and intuitive manner. Users may just follow the auditory cues rather than continually looking at a device or map, allowing them to concentrate on their surroundings and enjoy the experience. Here are some of the reasons why spatial audio is crucial in augmented reality: \textbf{Realism:} By replicating how sound interacts in the actual environment, spatial audio may assist in creating a more realistic audio experience in AR. Spatial audio may make digital music appear to come from a real-world item or location by simulating how sound waves reflect, reverberate, and move across space. \textbf{Contextualization}: Spatial audio may also give the user contextual information. It can, for example, be used to highlight certain items in the environment or to lead people to specific areas using aural cues.
\textbf{Immersion}: By giving auditory feedback timed with their motions and activities in the actual environment, spatial audio may make users feel like they are truly a part of the augmented reality experience.
\textbf{Accessibility}: Those with hearing difficulties may also benefit from spatial audio. Spatial audio can assist these people in better comprehending and participating in the augmented reality experience by giving auditory cues linked to visual information.
To summarize, spatial audio is critical for producing an engaging and immersive AR experience. It can enhance the user experience by providing realism, contextualization, immersion, and accessibility. By offering extra sensory clues, lowering cognitive burden, and making interaction with the environment more natural and intuitive, spatial auditory navigation can improve the overall user experience during augmented outdoor navigation activities.

\section*{Methods}
\label{sec:methods}
To explore the impact of spatial auditory navigation on user experience in an outdoor AR environment, a mobile AR application was designed based on the tourism needs in \textit{Tiergarten, Berlin}. The app prototype was implemented via \textit{Unity} program with an Android AR setup. The \textit{FMOD} plugin \footnote[1]{https://www.fmod.com/unity} was used in \textit{Unity} for setting the 3D sound; the \textit{Mapbox} SDK \footnote[2]{https://www.mapbox.com/} was used for setting GPS in the system and detecting the device’s location. The application aimed to guide tourists to reach out to several statues spread over the park through the navigation system and allowed them to explore the selected statues by providing information with an augmented presentation. Two conditions were designed with navigation modality as the control variable to compare the effects of \emph{spatial auditory navigation} and \emph{augmented visual navigation}. 

In the \emph{spatial auditory navigation} condition, users are navigated by spatial audio to find the “\textit{Goethe Monument.”} Technically, the audio source is placed at the GPS position of \textit{Goethe Monument}; with the 3D sound settlement in \textit{Unity}, end-users can judge direction and distance by variations in different sound attributes like timing, volume, etc. The 3D sound only appears when a user is in the activation zone with a radius value of 300 meters. The system gets the GPS information of the user's device in real-time; when the system detects the collision between the device and the destination, it jumps to the exploration page automatically, so the user knows the successful reach out. 

In the \emph{augmented visual navigation} condition, augmented visual cues are used on the device screen to navigate users to the \textit{“Beethoven-Haydn-Mozart Memorial.”} In this case, end-users can see a line directly pointing to the destination, and a floating board shows the distance between the user’s device and the destination. Same as for the \emph{spatial auditory navigation} condition, the page transition happens when the GPS position collision is detected. 

The experiment took place at \textit{Tiergarten} park, an outdoor environment, and all participants were asked to start from the same point. During the experiment, every participant went through a series of tasks that included both navigation conditions (\emph{spatial auditory navigation} and \emph{augmented visual navigation}) one by one with the task of finding specific statues and filling in questionnaires right after each task. 

The experiment was in a within-subjects design; in total, 20 participants (9 male and 11 female) were recruited and finished the experiment. 10 participants ranged in age of 18-34, 10 participants ranged in age of 25-54. Regarding working status, 60\% of the participants were students and 40\% were non-students. The participants were from different countries, with 12 from Germany, and 8 from other countries. The implementation of the experiment was approved by the Ethics Committee of \textit{Faculty IV} of \textit{Technische Universität Berlin}, all the participants have signed the consent form. First, the moderator gave participants a brief introduction to the application, and a mobile Android phone and headphones were provided. After filling in the demographic questionnaire and Affinity for Technology Interaction (ATI) Scale questionnaire \cite{ATI}, participants launched the app and read through the instruction. Then, they clicked the “Start” button and entered the selection page to choose a target destination. Because of the limited capabilities of the prototype, they were asked to choose \textit{“Goethe Monument”} as the first site and, based on their directional and distance sensation to the spatial music to reach the site, the music was a piece of guitar playing minuet in c in the classical style. Right after the session, they were asked to fill in a short version of the User Experience Questionnaire (UEQ-S) \cite{UEQS} to evaluate the experience with spatial auditory navigation. Next, participants went to the selection page and were asked to visit \textit{“Beethoven Memorial”} as the second site. Following the visual AR navigation cues on the screen, they arrived at the second site. \textit{Beethoven’s} work \textit{"Adieu au Piano"} was played in 2D through headphones during this session. Afterward, they filled in the UEQ-S for visual AR navigation experience evaluation. In the end, some pre-formulated open questions were asked to collect qualitative data and discuss the results.

\section*{Results}
\label{sec:Wichtig} 
\begin{figure}[h]
    \centering
    \includegraphics[width=0.5\textwidth]{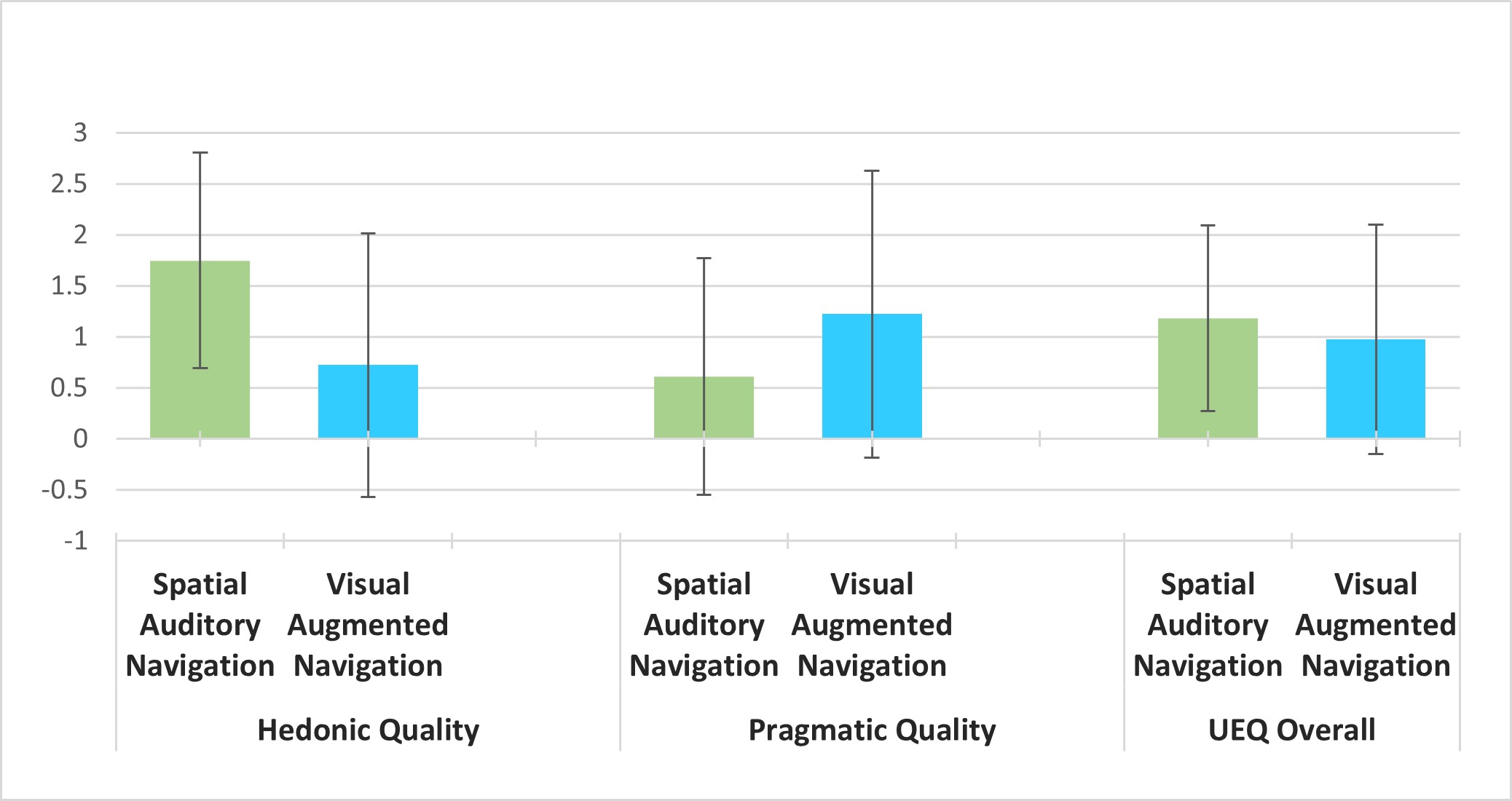}
    \caption{Mean of Hedonic Quality, Pragmatic Quality, UEQ Overall by Navigation Modality}
    \label{fig:mean1}
\end{figure}

The average Affinity for Technology Interaction of the participants was 3.86 (SD = 0.75). The data collected from the experiment was entered into UEQ Data Analysis Tool. A Paired-samples T-Test was conducted to determine statistically significant differences in Pragmatic Quality and Hedonic Quality between the two navigation modalities (see Figure \ref{fig:mean1}). 

The results showed that \emph{spatial auditory navigation} \textit{(M = 1.750; SD = 1.058)} performed significantly better \textit{(t (19) = 2.983; p = 0.008)} than \emph{augmented visual navigation} \textit{(M = 0.725; SD = 1.292)} on Hedonic Quality score during the outdoor environment. In detail, the main effect of navigation modalities on user experience Hedonic Quality revealed that spatial auditory navigation is more exciting, engaging, and inventive. In terms of the Pragmatic Quality score, \emph{spatial auditory navigation} \textit{(M = 0.613; SD = 1.163)} performed lower than \emph{augmented visual navigation} \textit{(M = 1.225; SD = 1.405)}; the results showed a trend toward a significant difference between the two navigation modalities \textit{(t (19) = 2.983; p = 0.089)} but did not reach statistical significance. When it comes to the UEQ Overall, the \emph{spatial auditory navigation} \textit{(M = 1.183; SD = 0.909)} performed better than \emph{augmented visual navigation} \textit{(M = 0.976; SD = 1.122)}, the result \textit{(t (19) = 0.801; p = 0.433)} indicated that while there is a difference in the UEQ Overall score between the two groups, it is not strong enough to be considered significant. More research is needed to confirm these findings. 

Furthermore, when comparing the results with the UEQ benchmark, the \emph{spatial auditory navigation} received a UEQ-Overall mean score of 1.18, which is above average compared to the benchmark, especially the Hedonic Quality achieved an “excellent” ranking. However, \emph{augmented visual navigation} received an UEQ-Overall score of 0.98, which is below average compared to the benchmark.

\section*{Discussion}
\label{sec:discussion} 
This work aimed to look at the influence of spatial auditory navigation on user experience in an outdoor environment using a mobile AR application meant to assist tourists. The research included two conditions with navigation modality as the control variable to examine the effects of spatial auditory navigation and augmented visual navigation.

By providing extra sensory signals that complement or replace visual clues, spatial auditory navigation may improve the user experience in augmented outdoor navigation activities. It gives users more information about their surroundings, such as the location and distance of items, landmarks, and areas of interest. Spatial audio is important in AR because it delivers a more realistic auditory experience, contextual information, immersion, and accessibility. In principle, spatial audio is required to produce an engaging and immersive AR experience, improving the user experience during augmented outdoor navigation activities.

The \emph{spatial auditory navigation} condition was shown to have a higher Hedonic Quality score than the \emph{augmented visual navigation} one. This result is probably related to some participants reporting that spatial auditory navigation was novel and exciting, letting them be more present and enjoy the environment more. 

The \emph{spatial auditory navigation} condition got a lower Pragmatic quality score than the \emph{augmented visual navigation}.
Despite not reaching a statistically significant difference, this outcome seems to come from users’ perception of \emph{spatial auditory navigation} condition, which was reported as more inaccurate in providing precise distance, orientation, and direction information than the augmented visual navigation one. Therefore, \emph{augmented visual navigation} results in more efficiency in completing a navigation task.

\bibliographystyle{plain}
\bibliography{main.bib}

\end{document}